\documentclass[aip,apl,reprint]{revtex4-1}

\usepackage[utf8]{inputenc}
\usepackage{amsmath}
\usepackage{amssymb}
\usepackage{amsfonts}
\usepackage{graphicx}
\usepackage{siunitx}					
\usepackage{color}

\begin{document}

\title{``Noiseless'' thermal noise measurement of atomic force microscopy cantilevers}

\author{Basile Pottier}
\affiliation{Univ Lyon, Ens de Lyon, Univ Claude Bernard Lyon 1, CNRS, Laboratoire de Physique, F-69342 Lyon, France}

\author{Ludovic Bellon}
\email{ludovic.bellon@ens-lyon.fr}
\affiliation{Univ Lyon, Ens de Lyon, Univ Claude Bernard Lyon 1, CNRS, Laboratoire de Physique, F-69342 Lyon, France}

\date{\today}

\begin{abstract}
When measuring quadratic values representative of random fluctuations, such as the thermal noise of Atomic Force Microscopy (AFM) cantilevers, the background measurement noise cannot be averaged to zero. We present a signal processing method that allows to get rid of this limitation using the ubiquitous optical beam deflection sensor of standard AFMs. We demonstrate a two orders of magnitude enhancement of the signal to noise ratio in our experiment, allowing the calibration of stiff cantilevers or easy identification of higher order modes from thermal noise measurements.
\end{abstract}

\maketitle 

A common procedure to calibrate the stiffness of Atomic Force Microscopy (AFM) cantilevers uses the thermal noise driven fluctuations of its deflection~\cite{hutter_calibration_1993,Butt-1995}. Implemented in most commercial devices, this method consists in measuring the mean-square amplitude of the cantilever's deflection $\langle x^2 \rangle$ at thermal equilibrium around $\langle x \rangle=0$. The equipartition principle states that the energy stored in the spring is in average equal to the thermal energy thus
\begin{equation}\label{equipartition}
	\frac{1}{2}k\langle x^2 \rangle=\frac{1}{2}k_BT
\end{equation}
where $k_B$ (the Boltzmann's constant) and $T$ (the temperature) are known, and $k$ is the spring constant to calibrate. At room temperature, these tiny mechanical fluctuations range from $\SI{200}{pm\ rms}$ for a soft probe ($\SI{0.1}{N/m}$) to only $\SI{10}{pm\ rms}$ for a stiff probe ($\SI{40}{N/m}$). One is thus very sensitive to presence of the noise in the measurement. Indeed, let
\begin{equation}
X=x+\eta
\end{equation}
be the measured value of the deflection $x$ and $\eta$ the measurement noise, the contribution of the latter doesn't vanish when computing mean square quantities even after a long averaging:
\begin{equation}
	\langle X^2 \rangle = \langle x^2 \rangle + \langle \eta^2 \rangle = \frac{k_BT}{k} + \langle \eta^2 \rangle
\end{equation}
where the cross term $2 \langle x \eta \rangle= 2 \langle x \rangle\langle \eta \rangle$ averages to 0 since the two stochastic signals are uncorrelated and of zero mean. The stiffer the cantilever, the smaller is the signal to noise ratio $\langle x^2 \rangle/\langle \eta^2 \rangle$ and the less accurate is the spring constant calibration.

In order to isolate the contribution of the mechanical fluctuations, one studies the Power Spectral Density (PSD) $S_{X^{2}}$ of the signal $X$:
\begin{equation}
S_{X^{2}}(f)=\frac{\langle|\tilde{X}(f)|^{2}\rangle}{\Delta f}
\end{equation}
where $\tilde{X}(f)$ is the Fourier transform of $X$ and $\Delta f$ the spectral bins bandwidth. Indeed, thanks to the resonant behavior of the cantilever near its normal modes, the energy of mechanical fluctuations is mainly stored around the resonance frequencies (especially the first one)~\cite{Butt-1995}. The measurement noise, on the contrary, is often only a white noise, its energy being flat across all the frequency spectrum. This is the case for instance for the shot noise of an optical measurement. It is thus convenient to study the first resonance of the measured spectrum $S_{X^{2}}$, which can be fitted by the prediction of the Fluctuation-Dissipation Theorem (FDT) for a Simple Harmonic Oscillator (SHO), with an additional white noise $S_{\eta^{2}}$:
\begin{align}
	S_{X^{2}}(f) &= S_{x^{2}}(f) + S_{\eta^{2}}\\
	&=  \frac{2k_BT}{\pi k f_0} \frac{1/Q}{\left( 1-u^2 \right)^2+u^2/Q^2} + S_{\eta^{2}} \label{EqSHO}
\end{align}
where $Q$ is the quality factor and $u=f/f_0$ is the frequency normalized to the resonance one $f_0$. The equipartition theorem (\ref{equipartition}) can be recovered from the FDT by integrating this last equation over all frequencies, but thanks to the resonance the signal to noise ratio is increased by a factor $Q$ near $f_{0}$. This strategy is efficient and can give reasonable results for stiffnesses up to $\SI{1}{N/m}$, but uncertainty is large for stiffer cantilevers, for which  the mechanical fluctuation $S_{x^{2}}$ only stand above the background noise $S_{\eta^{2}}$ in a tiny frequency range around the resonance. Minimizing the $S_{\eta^{2}}$ contribution to the PSD is thus highly desirable.

To get rid of this problem, let us perform two simultaneous independent measurements $X_{1}$ and $X_{2}$ of the deflection $x$, each being plagued by an independent noise $\eta_{1}$ and $\eta_{2}$:
\begin{equation}
X_{1} =x+\eta_{1} , \quad X_{2}=x+\eta_{2}
\end{equation}
The Cross Spectrum Density (CSD) $S_{X_{1}\!X_{2}}$ of these two signal then directly averages to the PSD of interest\cite{Mitsui-2009,Mitsui-2013} $S_{x^{2}}$:
\begin{equation} \label{eq:cross}
S_{X_{1}\!X_{2}} (f)=\frac{\langle \tilde{X_{1}}(f)\tilde{X_{2}^{*}}(f)\rangle}{\Delta f} = S_{x^{2}}(f)
\end{equation}
Indeed, all cross spectra corresponding to uncorrelated stochastic signals ($\langle x \eta_{1} \rangle$, $\langle x \eta_{2} \rangle$, $\langle \eta_{1} \eta_{2} \rangle$) average to 0. This simple operation thus allows the complete canceling of the measurement noise, as long as it is uncorrelated to the mechanical fluctuations of the cantilever. In this letter, we present an implementation of this technique using the ubiquitous optical beam deflection (OBD) technique.

\begin{figure}[!ht]
\begin{center}
 \includegraphics[width=6cm]{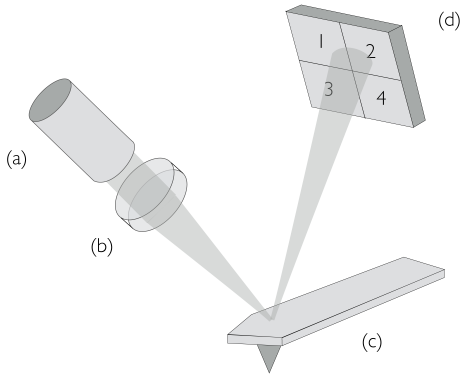}
 \caption{\label{Fig1} Schematic representation of the optical beam deflection technique: the beam of the laser (a) is focused with lens (b) on the free end of the cantilever (c). The reflected beam is centered on the four quadrants photodiode (d). Its position is a function of the local slope of the cantilever. The fluctuations of the beam position (measured through the photocurrents delivered by the quadrants) therefore provide a measurement of the fluctuations of the cantilever flexion and torsion.}
 \end{center}
\end{figure}

Let us first review the classic OBD method and its limitations. Introduced by Meyer and Amer~\cite{meyer_novel_1988}, this technique is the most widely used in laboratory and commercially available AFMs: it combines ease of use and high resolution. It consists in illuminating the free end of the cantilever with a focused laser beam and detecting the fluctuations of the reflected beam position with a 4-quadrants photodiode (figure~\ref{Fig1}). The reflected beam position is function of the local slope of cantilever free end, thus representative of its flexion (angle $\theta$) and torsion ($\theta'$). Practically the position of reflected beam is measured by computing a contrast function $C_0$ of the photocurrents $I_n$ delivered by the 4 quadrants $n=1$ to $4$ of the photodiode:
\begin{equation} \label{eq:C0}
	C_{0}=\frac{I_{1}+I_{2}-I_{3}-I_{4}}{I_{1}+I_{2}+I_{3}+I_{4}}
\end{equation}
A downward flexion of the cantilever will move the laser spot towards the top quadrants (1,2), increasing $I_{1}$ and $I_{2}$ and decreasing $I_{3}$ and $I_{4}$, thus $C_{0}$ is a measurement of the flexion angle $\theta$, hence of the deflection $x$. A similar contrast $C_{0}'$ between left (1,3) and right (2,4) quadrants intensities gives a measurement of the torsion angle $\theta'$.

At first order, $C_{0}$ is proportional to $\theta$ or $x$:
\begin{equation}
	C_{0}=\sigma_{\theta} \theta + \eta_{0}=\sigma_{x} x + \eta_{0}
\end{equation}
with $\eta_{0}$ the measurement noise, and $\sigma_{\theta}$ or $\sigma_{x}$ the sensitivity of the sensor. The thermal fluctuations spectra $S_{x^{2}}$ or $S_{\theta^{2}}$ will thus be hidden when the amplitude of the noise is too high:
\begin{equation}
S_{C_{0}^{2}}=\sigma_{\theta}^{2} S_{\theta^{2}}+S_{\eta_{0}^{2}}=\sigma_{x}^{2} S_{x^{2}}+S_{\eta_{0}^{2}}
\end{equation}
Using gaussian beam optics, the theoretical value of the angular sensitivity can be computed~\cite{Paolino-2007}:
\begin{equation}
	\sigma_{\theta} = \sqrt{8\pi} \frac{R}{\lambda}
\end{equation}
where $\lambda$ is the laser wavelength and $R$ the radius of the beam waist. The deflection sensitivity $\sigma_{x}$ depends on the cantilever length $L$ and the oscillation mode $m$ of the cantilever: using an Euler-Bernoulli description of a rectangular cantilever~\cite{Butt-1995}, it can be shown that $\sigma_{x}= \beta_{m} \sigma_{\theta}/L$, with $\beta_{1}\approx 1.38$, $\beta_{2}\approx 4.78$, \ldots $\beta_{m}\approx (m-1/2)\pi$. In practice, $\sigma_{x}$ is approximately  calibrated for mode 1 using a force curve on a hard surface (usually $\sigma_{x}$ is given in $\SI{}{nm/V}$ in commercial AFMs). 

The source of noises contributing to $\eta_{0}$ can be classified into two categories: classical noise, like the light source having fluctuation in intensity or beam direction, and quantum noise, like the photon shot noise at the detector and the back-action of the light on the cantilever~\cite{smith_limits_1995}. The effects of classical noise can be reduced by proper experimental design: the effect of fluctuations in light intensity for example are canceled using the contrast $C_{0}$ where the difference between the intensities on the top quadrants and the bottoms ones is normalized by the total intensity (eq. \ref{eq:C0}). On the other hand, quantum mechanical noises are unavoidable. The optical shot noise, which is due to the counting statistics of the photons, is in most cases the dominant source of detection noise and limits the resolution of the OBD technique. According to the Poisson distribution of the light source, the shot noise is frequency independent and appears as a white noise in all measured intensities $I_{n}$: $S_{I_{n}^{2}}^{\mathrm{SN}}=2eI_{n}$, with $e$ the elementary charge. Using the definition of $C_{0}$ (eq. \ref{eq:C0}), assuming the beam is centered on the photodiode, we compute the flat baseline which represents the shot noise limit for the deflection measurement:
\begin{equation} \label{eq:SN}
S_{\eta_{0}^{2}}^{\mathrm{SN}}=\sum_{n=1}^{4} \left(\frac{\partial C_{0}}{\partial I_{n}}\right)^{2} S_{I_{n}^{2}}^{\mathrm{SN}} = \frac{2 e}{I_{0}}
\end{equation}
where $I_{0}=\sum_{n=1}^{4} I_{n}$ is the total photocurrent. Other contributions may sum on top of this shot noise limit, such as the ones from the conditioning electronics (Johnson-Nyquist noise of the current amplifier (white noise), $1/f$ noise, high frequency electronics noise, etc.), from the analog to digital convertion\ldots\ One might notice that we can reduce the shot noise level by increasing the light beam power. However in practice the laser power is limited. Besides, because of partial light absorption by the cantilever, raising the power risks to significantly heat the cantilever, which may alter its stiffness or functionality \cite{sandoval_resonance_2015}. Typical shot noise limit is $S_{\eta_{0}^{2}}^{\mathrm{SN}}/\sigma_{x}^{2}\approx\SI{4E-26}{m^{2}/Hz}$ (numerical application for a $L=\SI{125}{\micro m}$ long cantilever, using a $\SI{0.1}{mW}$ laser beam at $\lambda=\SI{633}{nm}$ focused on a $R=\SI{5}{\micro m}$ radius spot, with a photodiode efficiency of $\SI{0.4}{A/W}$). This is close to the thermal fluctuations expected at resonance for a stiff cantilever: $S_{x^{2}}(f_{0})\approx\SI{9E-26}{m^{2}/Hz}$ for $f_{0}=\SI{300}{kHz}$, $Q=400$, $k=\SI{40}{N/m}$ at room temperature. Note that for AFM experiments where the cantilever is immersed in water rather than in air (corresponding to a smaller quality factor $Q$), the viscosity of the environment broadens the structural resonances and decreases the peak amplitude. In these cases, the calibration procedure is restricted to even less stiff cantilevers. 
  
We should note that the usage of nonclassical light technique (e.g. squeezed states of light~\cite{caves_quantum-mechanical_1981,fabre_quantum_2000}) provides ways of decreasing the noise floor of OBD technique below the shot noise limit~\cite{treps_surpassing_2002}. Nevertheless these quantum optics techniques involve exotic experimental setup and the gained factor on the signal-to-noise ratio is at most ten~\cite{pooser_ultrasensitive_2015, vahlbruch_observation_2008}. In the next paragraph we show that our classical technique, easy to implement, allows to statistically average to zero the shot noise contribution in the measured power spectrum. 

The drawback of the cross correlation approach presented in the introduction (eq. \ref{eq:cross}) is the need of two independent measurement of the deflection, apparently implying a second sensor~\cite{Mitsui-2009,Mitsui-2013}. We propose here a patented method\cite{Bellon-2015-CrossCor} using only the common 4-quadrants photodetector (figure~\ref{Fig1}). Indeed, this sensor directly gives access to two independent measurements of the deflection. We define for this purpose two contrasts
\begin{equation}
	C_1=\frac{I_1-I_3}{I_1+I_3}, \quad  C_2=\frac{I_2-I_4}{I_2+I_4}
\end{equation}
Those signals are equivalent to the global contrast $C_{0}$, in the sense that they are each a direct measurement of the angular deflection $\theta$ (or vertical deflection $x$) of the cantilever: $C_{n}=\sigma_{\theta} \theta + \eta_{n}=\sigma_{x} x + \eta_{n}$ for $n=0$ to $2$. Since each contrast $C_{1}$ and $C_{2}$ involves an independent pair of quadrants, the detection noises $\eta_{1}$ and $\eta_{2}$ associated to the two signals are independent. The CSD will then converge to the PSD of thermal fluctuations of the cantilever:
\begin{equation}
S_{C_{1}C_{2}}=\sigma_{\theta}^{2} S_{\theta^{2}}=\sigma_{x}^{2} S_{x^{2}}
\end{equation}
This method is very simple to implement, as it does not imply any modification of conventional AFM setups, except for the data acquisition and processing. Note that torsion has no effect on $C_{1}$ and $C_{2}$, as the normalisation per pair cancels any lateral positioning effect: if the laser beam move leftwards for instance, both $I_{1}$ and $I_{3}$ increase, but their ratio (thus $C_{1}$) stay unchanged. Swapping the role of quadrant 2 and 3 in the formulas, one can equivalently define two signals only sensitive to torsion. 

\begin{figure}[!ht]
\begin{center}
\includegraphics{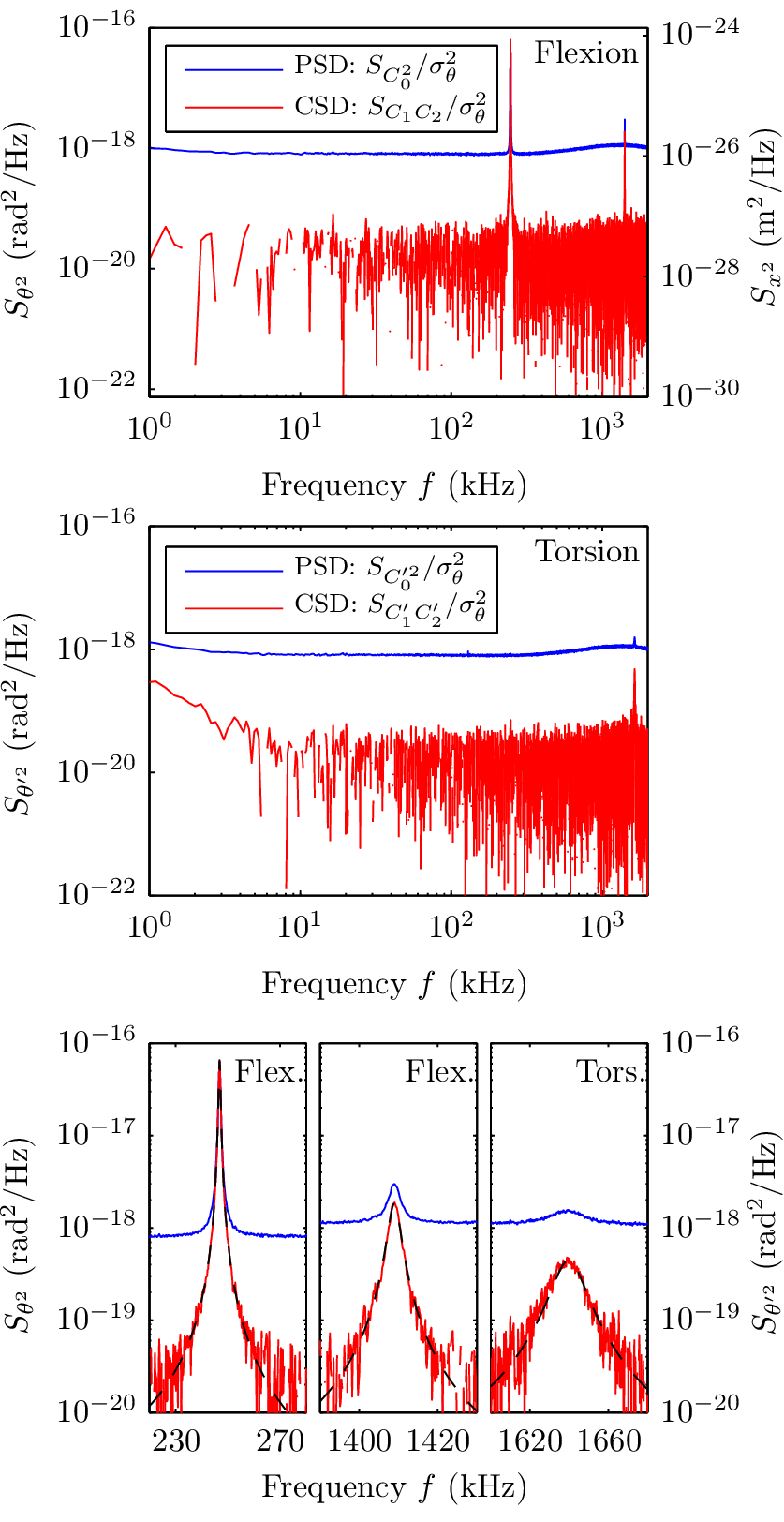}
\caption{\label{Fig2}(Color online) Thermal noise spectra on a stiff cantilever. The 2 top graphs present the thermal noise spectra measured in flexion and torsion on the full frequency range from a $\SI{10}{s}$ data acquisition, and the 3 bottom graphs are zooms on the resonances (2 in flexion, 1 in torsion). On the top graph, the $y$ scale is dual: angular deflection can be read on the left scale, vertical deflection on the right scale (note that the right scale calibration is only valid for mode 1). On each graph, the PSD (blue) is shot noise limited: a noise floor at $\SI{E-18}{rad^{2}/Hz}$ hides any features of the thermal fluctuations of the cantilever below this level. This shot noise contribution averages toward zero for the CSD (red): the rms value of the remaining noise is 40 times smaller after $\SI{10}{s}$ averaging. The black dashed curves are fits of the resonances with the prediction of the FTD for a simple harmonic oscillator.}
\end{center}
\end{figure}

In order to illustrate the effectiveness of the noise reduction method, we compare the thermal fluctuation spectra of a cantilever computed in the usual way (with $S_{C_{0}^{2}}$) and using the cross correlation procedure ($S_{C_{1}C_{2}}$) on the same acquisition data. The measurements are performed in air at room temperature ($T=\SI{295}{K}$) on a commercial silicon cantilever with aluminium reflex coating (Olympus AC160TS-R3). Its nominal characteristics are: length $L=\SI{160}{\micro m}$, stiffness $k=\SI{26}{N/m}$, resonant frequency $f_{0}=\SI{300}{kHz}$. We use a home built OBD setup, with a stabilized solid state laser from Spectra Physics ($\SI{1}{mW}$ at $\lambda=\SI{532}{nm}$). The laser beam is focused on the tip with a $\SI{30}{mm}$ achromatic lens, resulting in a diffraction limited circular spot (measured radius: $R=\SI{5.4}{\micro m}$). After reflection, the laser beam is sent at the center of a 4 quadrants photodiode (SPOT-4D from Osioptoelectronics).  The photocurrents delivered by each quadrant of the segmented photodiode are converted to voltages by 4 independent home made low noise preamplifiers (AD8067 operational amplifiers with $\SI{100}{k \ohm}$ retroaction and $\SI{1}{MHz}$ bandwidth). Analog to digital conversion is performed by NI-PXI 5922 digitizers (ADC, 18 bits, sampling rate $\SI{6}{MHz}$, National Instruments). Therefore we measure the four photocurrents $I_{n}$, and using post-acquisition signal processing, we compute numerically the contrasts $C_{0}$, $C_{1}$ and $C_{2}$ for the flexion, and their equivalent for torsion. We then compute the thermal noise through the PSD of $C_{0}$, and through the CSD of $C_{1}$ and $C_{2}$, both in flexion and torsion. Each PSD or CSD is computed using a Welch averaging on the same given time window ($T=\SI{10}{s}$), using $2^{15}$ points for the Fourier transform. This results in a $\Delta f=\SI{183}{Hz}$ spectral resolution. We use the theoretical value for the sensitivity $\sigma_{x}$ or $\sigma_{\theta}$. Note that $\sigma_{x}$ is mode dependent, we use here only the value of mode 1: the scale for $S_{x^{2}}$, in $\SI{}{m^{2}/Hz}$, is thus only meaningful for the first mode in figure~\ref{Fig2}. The scale for $S_{\theta^{2}}$, in $\SI{}{rad^{2}/Hz}$, is accurate on all the frequency range.

Figure~\ref{Fig2} displays the measured spectra, with a zoom on the two first modes in flexion and the first mode in torsion that are visible in the explored frequency range. The background noise of the PSD is mainly due to the shot noise of the photodiodes: it is frequency-independent and its magnitude corresponds to the theoretical amplitude calculated according to Eq.~(\ref{eq:SN}), with $I_{0}=\SI{164}{\micro A}$ for this specific measurement. Above $\SI{200}{kHz}$, the experimental noise of the PSD deviates slightly from the expected white noise due to an additional contribution from the conditioning electronics.
We can observe on the CSD that using the cross correlation approach, the background noise is decreased with respect to the PSD by a factor of around 40 after only 10 seconds of averaging. In fact, when the cantilever thermal noise is not dominant, the amplitude of the CSD tends to zero in average, and we only see the positive values in the displayed spectrum due to the log scale (the CSD is complex, we display only its real part which can be negative). This is also the reason why there are some blanks in the CSD curve far away from the resonances. The thermal fluctuations of the cantilever are much easier to see on the CSD than on the PSD: the torsional mode for example presents a signal to noise ratio below one with the PSD, but clearly emerges from the background noise in the CSD.

\begin{figure}[t]
\begin{center}
\includegraphics{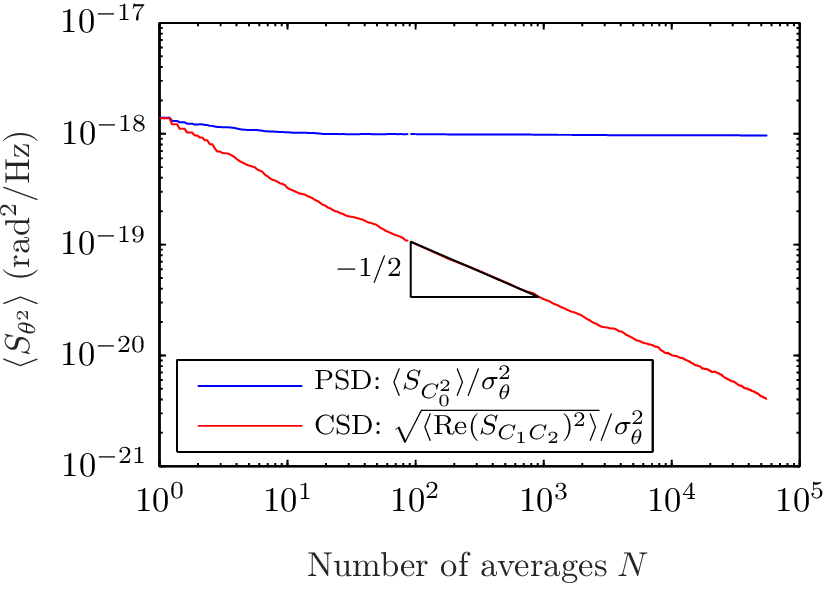}
 \caption{\label{Fig3}Mean value of the PSD and rms value of the real part of the CSD against the number of averages $N$. Each value is computed from the spectrum in flexion in the frequency range $\SI{650}{kHz}<f<\SI{750}{kHz}$. While the PSD directly converges towards a background noise, the CSD starts from the same level but the rms of the remaining noise decreases as $1/\sqrt{N}$.}
\end{center}
\end{figure}

Let us now discuss the efficiency of the cross-correlation technique. In every frequency bin of the CSD, the root mean square value (rms) of the shot noise contribution during the averaging process decays as $1/ \sqrt{N}$, where $N$ is the number of averages. This number of averages $N$ depends on the frequency resolution $\Delta f$ and acquisition length $T$: $N=T \Delta f$. The reduction  of noise that can be reached depends on the properties of the signal. Here with $\Delta f=\SI{183}{Hz}$ and $T=\SI{10}{s}$ , we have for instance N=1830. So the apparent noise in $\SI{}{m^{2}/Hz}$ or $\SI{}{rad^{2}/Hz}$ is reduced by a factor $\sqrt{1830}\approx 42$. In order to verify the decay rate $1/ \sqrt{N}$ of the remaining rms noise, we plot in figure~\ref{Fig3} the mean value of the PSD and the root mean square value of the CSD for the measurement in flexion around $\SI{700}{kHz}$ (where only measurement noise is expected) as a function of the number of averages $N$. While the PSD converges towards the background level of $\SI{E-18}{rad^{2}/Hz}$, the CSD starts from the same background noise but decreases as $1/\sqrt{N}$. At the last point displayed $N=\num{5.5E4}$, corresponding to 5 minutes of acquisition, the cross-correlation gained a factor of around 237 when we expect $\sqrt{N}=234$.

In conclusion, the presented method allows a significant reduction of the background noise in the measurement of thermal fluctuations of an AFM cantilever. It is easy to implement as it consists only in a signal processing technique applied to the the common opto-mechanical setup used in most AFMs. The only requirement is the direct access to the signals of the 4 quadrants of the photodiode. It can be used to enhance the signal to noise ratio, noteworthily for the calibration of stiff cantilevers, or for higher order modes identification. This simple method can be directly applied to other research areas where fluctuations are measured with a 4 quadrants photodiode, such as passive rheology~\cite{Mitsui-2009,Mitsui-2013,Pottier-2011,Pottier-2013,Pottier-2015-CrossCor}.

\begin{acknowledgments}
We thank F. Ropars and F. Vittoz for technical support, C. Fretigny, L. Talini, A. Petrosyan, S. Ciliberto for stimulating discussions. This work has been supported by the ANR project \emph{HiResAFM} (ANR-11-JS04-012-01) of the Agence Nationale de la Recherche in France.
\end{acknowledgments} 

%\bibliographystyle{unsrt}
%\bibliography{ArticleCross}

\end{document}